\documentstyle[aps,prl,preprint,floats,epsfig]{revtex}

\begin{document}

\preprint{\tighten\vbox{\hbox{\hfil CLNS 01/1741}
                        \hbox{\hfil CLEO 01-13}
}}

\title{First Measurement of $\Gamma(D^{\ast +})$ and\\
       Precision Measurement of $m_{D^{\ast +}} - m_{D^0}$ }

\author{CLEO Collaboration}
\date{8 August 2001}

\maketitle
\tighten

\begin{abstract} 
We present the first measurement of the $D^{\ast +}$ width using
9/fb of $e^+e^-$ data collected near the $\Upsilon(4\rm{S})$
resonance by the CLEO II.V detector.
Our method uses advanced tracking techniques
and a reconstruction method that takes advantage of the small
vertical size of the CESR beam spot to measure the energy release
distribution from the $D^{\ast +} \to D^0 \pi^+$ decay.
We find $\Gamma(D^{\ast +}) = 96 \pm 4\ ({\rm Statistical})
\pm 22\ ({\rm Systematic})$ keV.  We also measure the energy
release in the decay and compute $\Delta m \equiv 
m_{D^{\ast +}} - m_{D^0} = 145.412 \pm 0.002 ({\rm Statistical})
\pm 0.012\ ({\rm Systematic})$ MeV/c$^2$.

\end{abstract}
\newpage

{
\renewcommand{\thefootnote}{\fnsymbol{footnote}}

\begin{center}
A.~Anastassov,$^{1}$ E.~Eckhart,$^{1}$ K.~K.~Gan,$^{1}$
C.~Gwon,$^{1}$ T.~Hart,$^{1}$ K.~Honscheid,$^{1}$
D.~Hufnagel,$^{1}$ H.~Kagan,$^{1}$ R.~Kass,$^{1}$
T.~K.~Pedlar,$^{1}$
% ...
\end{center}
 
\small
\begin{center}
$^{1}${Ohio State University, Columbus, Ohio 43210}\\
$^{2}${University of Oklahoma, Norman, Oklahoma 73019}\\
% ...
\end{center}

\setcounter{footnote}{0}
}
\newpage

 
\section{Introduction}

A measurement of $\Gamma(D^{\ast +})$ opens an important window
on the non-perturbative strong physics involving heavy quarks.
The basic framework of the theory is well understood, however, there is
still much speculation -
predictions for the width range from $15\,\rm{keV}$ to 
$150\,\rm{keV}$ \cite{pred}.
The level splitting in the $B$ sector
is not large enough to allow real strong transitions.
Therefore, a measurement of the width of the $D^{\ast +}$
gives unique information about the
strong coupling constant in heavy-light meson systems.

	The total width of the $D^{\ast +}$ is the sum of the partial widths
of the strong decays $D^{\ast +} \to D^0 \pi^+$ and $D^{\ast +} \to D^+ \pi^0$
and the electromagnetic decay $D^{\ast +} \to D^+ \gamma$.  We can write
the width in terms of strong couplings,
$g_{D^\ast \to D^0\pi^+}$ and $g_{D^\ast \to D^+\pi^0}$,
and an electromagnetic coupling, $g_{D^\ast \to D^+ \gamma}$:
\begin{eqnarray}
\Gamma(D^{\ast +}) & = & 
              \Gamma(D^0 \pi^+) + \Gamma(D^+ \pi^0) + \Gamma(D^+ \gamma) \\
                   & = & \frac{g_{D^\ast \to D^0\pi^+}^2}{24\pi m_{D^{\ast +}}^2}p^3_{\pi^+} +
                         \frac{ g_{D^\ast \to D^+\pi^0}^2}{24\pi m_{D^{\ast +}}^2}p^3_{\pi^0} +
                         \frac{\alpha g_{D^\ast \to D^+ \gamma}^2}{3}p^3_{\gamma},
\end{eqnarray}
where the momenta are those for the indicated particle in the
$D^{\ast +}$ rest frame, and $\alpha$ is the
fine structure constant.
This can be rewritten using the isospin relationship
\begin{equation}
g_{D^\ast \to D\pi} = -\sqrt{2} g_{D^\ast \to D^+ \pi^0} = g_{D^\ast \to D^0 \pi^+}, 
\end{equation}
and relating $g_{D^\ast \to D\pi}$ to a universal strong coupling between heavy
vector and pseudoscaler mesons to the pion, $g$, with
\begin{equation}
g_{D^\ast \to D\pi} = \frac{2 m_{D^{\ast}}}{f_\pi} g,
\end{equation}
where $f_\pi$ is the pion decay constant.
All this yields
\begin{equation}
\Gamma(D^{\ast +}) = \frac{2g^2}{12\pi f_\pi^2}p^3_{\pi^+} +
                     \frac{ g^2}{12\pi f_\pi^2}p^3_{\pi^0} +
                     \frac{\alpha g_{D^\ast \to D^+ \gamma}^2}{3}p^3_{\gamma}.
\end{equation}
The width of the $D^{\ast +}$ only depends on $g$~\cite{wise} since the
contribution of the electromagnetic decay
with branching fraction $(1.68\pm0.45)$\% \cite{mats} can be neglected.
The measurement of $g$ is needed in the extraction of $V_{ub}$
in semileptonic $b \to u$ decays~\cite{burdman}.

Prior to this measurement, the $D^{\ast +}$ width was limited
to be less than $131\,\rm{keV}$ at the $90\%$ confidence level 
by the ACCMOR collaboration \cite{ACCMOR}.
The limit was based on 110 signal events reconstructed in two
$D^0$ decay channels with a background of 15\%.
This contribution describes
a measurement of the $D^{\ast +}$ width with the CLEO II.V detector
where the signal, in excess of 11,000 events,  is
reconstructed through a single, well-measured sequence,
$D^{\ast +} \rightarrow \pi^+_{\rm slow} D^0$,
$D^0 \rightarrow K^-\pi^+$.  Consideration of
charge conjugated modes are implied throughout this paper.
The level of background under the signal is less than 3\%
in our loosest selection.

The challenge of measuring the width of the $D^{\ast +}$ is understanding the tracking
system response function since the experimental resolution exceeds the width
we are trying to measure. 
Candidates with tracks that have
 mismeasured hits, errors in pattern recognition, and
large angle Coloumb scattering are particularly dangerous because
the signal shape they project is broad and the errors for these events can be
underestimated, resulting in events that can easily influence the 
parameters of a Breit-Wigner fitting shape.  We generically term
such effects ``tracking mishaps.''
A difficulty is that there is no physical calibration for this measurement.
The ideal calibration mode would have a large
cross-section, a width of zero, decay with a rather small
energy release to three charged
particles one of which has a much softer momentum distribution than the
other two which decay through a nearly zero width resonance
with a measurable flight distance.
Such a mode would allow us to disentangle detector effects from the underlying
width but no such mode exists.

Therefore, to measure the width of the $D^{\ast +}$ we depend
on exhaustive comparisons between a GEANT \cite{GEANT} based detector
simulation and our data.  We addressed the problem by selecting samples of
candidate $D^{\ast +}$ decays using three strategies.

	First we produced the largest sample from data and simulation by 
imposing only
basic tracking consistency requirements.  We call this the
{\em nominal} sample.

	Second we refine the nominal sample selecting candidates
with the best measured tracks by making very tight cuts
on tracking parameters.  There is special emphasis on choosing those tracks
that are well measured in our silicon vertex detector.  This reduces
our nominal sample by a factor of thirty and, according to our
simulation, has negligible contribution from tracking mishaps.
We call this the {\em tracking selected} sample.

	A third alternative is to select our data on specific kinematic
properties of the $D^{\ast +}$ decay that minimize the dependence
of the width of the $D^{\ast +}$ on detector mismeasurements.
The nominal sample size is reduced by a factor
of three and a half and, again according to our simulation,
the effect of tracking problems is reduced to negligible levels.
We call this the {\em kinematic selected} sample.

	In all three samples the width is extracted with an unbinned maximum
likelihood fit to the energy release distribution and compared with
the simulation's generated value to determine a bias which is
then applied to the data.
These three different approaches yield consistent values for
the width of the $D^{\ast +}$ giving us confidence that our simulation
accurately models our data.


\section{CLEO Detector and Data Samples}

	The CLEO detector has been described in detail elsewhere.  All of
the data used in this analysis are taken with the detector in its
II.V configuration \cite{CLEO}.  This work mainly depends on the 
tracking system
of the detector which consists of a three-layer, double sided silicon
strip detector, an intermediate ten-layer drift chamber,
and a large 51-layer helium-propane drift chamber.  All three are in an
axial magnetic field of 1.5 Tesla provided by a superconducting solenoid
that contains the tracking region.  The charged tracks are fit using a
Kalman filter technique that takes into account energy loss as the tracks
pass through the material of the beam pipe and detector \cite{kalman}.

	The data were taken in symmetric $e^+e^-$ collisions at a center of
mass energy around 10 GeV with an integrated luminosity of 9.0/fb provided by
the Cornell Electron-positron Storage Ring (CESR).  The nominal sample
follows the selection of
$D^{\ast +} \to \pi_{\rm slow}^+ D^0 \to K^-\pi^+\pi_{\rm slow}^+$ candidates
used in our $D^0-\bar{D^0}$ mixing analysis\cite{Dmix}.

\begin{figure}
     \centerline{\epsfig{file=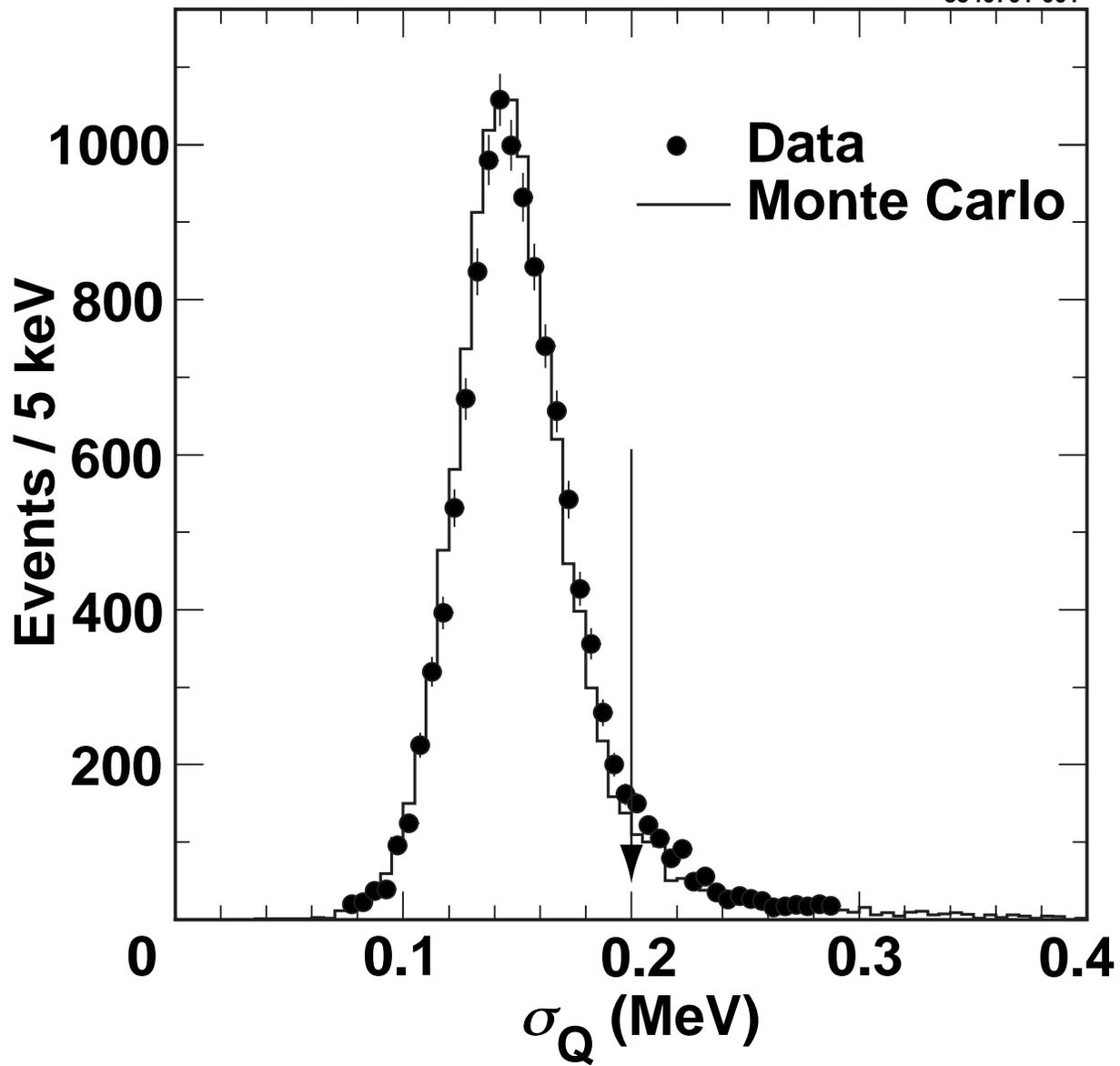,width=\linewidth}}
     \caption{\label{fig:errorcompare} Distribution of $\sigma_Q$,
              the uncertainty on $Q$ as determined from propagating
              track fitting errors.  The arrow indicates a selection
              discussed in the text.}
\end{figure}

	Our reconstruction method takes advantage of the small CESR beam
spot and the kinematics and topology of the
$D^{\ast +} \to \pi^+_{\rm slow} D^0 \to \pi^+_{\rm slow} K^- \pi^+$
decay chain. The $K^-$ and $\pi^+$ are required to form a common vertex.
The resultant $D^0$ candidate momentum vector is then projected back to the
CESR luminous region to determine the $D^0$ production point.  
The CESR luminous region has a Gaussian width $\sim 10\ \mu$m vertically and
$\sim 300\ \mu$m horizontally.  It is well determined
by an independent method\cite{hourglass}.
This procedure determines an accurate
$D^0$ production point for $D^0$'s moving out of the horizontal plane;
$D^0$'s moving within 0.3 radians of the horizontal plane are not
considered.  Then the $\pi_{\rm slow}^+$ track
is refit constraining its trajectory to intersect the $D^0$
production point.  This improves the resolution on the energy release,
$Q = M(K^-\pi^+\pi_{\rm slow}^+) - M(K^-\pi^+) - m_{\pi^+}$, by more
than 30\% over simply forming the appropriate invariant masses of the tracks.
The improvement to resolution is essential to our measurement of the width
of the $D^{\ast +}$.  Our resolution is shown in Figure~\ref{fig:errorcompare}
and is typically 150 keV.
The good agreement between 
Monte Carlo and data demonstrates that the kinematics and sources of 
uncertainties on the tracks, such as the number of hits used and the
effects of multiple scattering in detector material, are well modeled.

\begin{figure}
\begin{tabular}{cc}
\epsfxsize=85mm \epsfbox{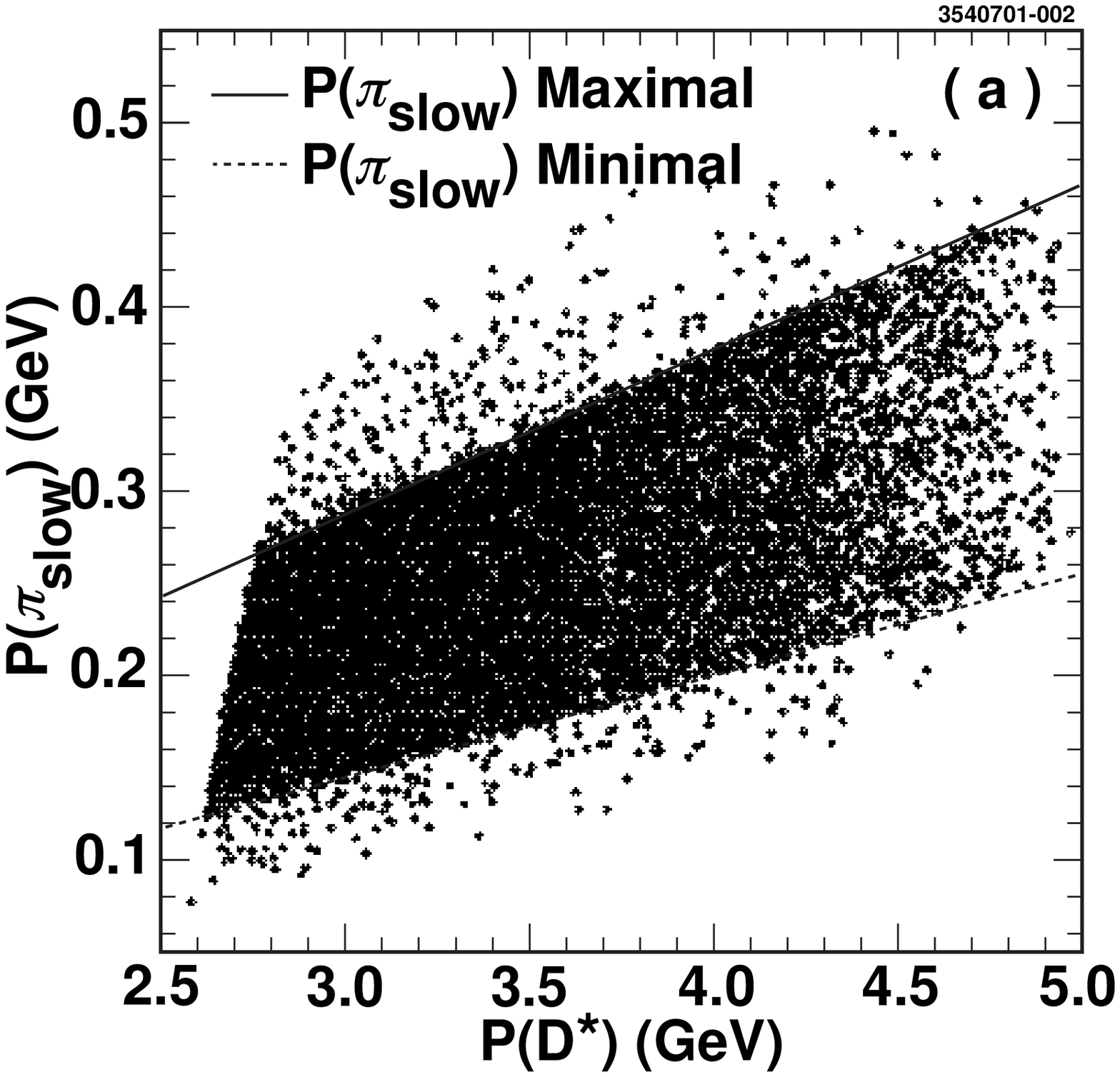} &
\epsfxsize=85mm \epsfbox{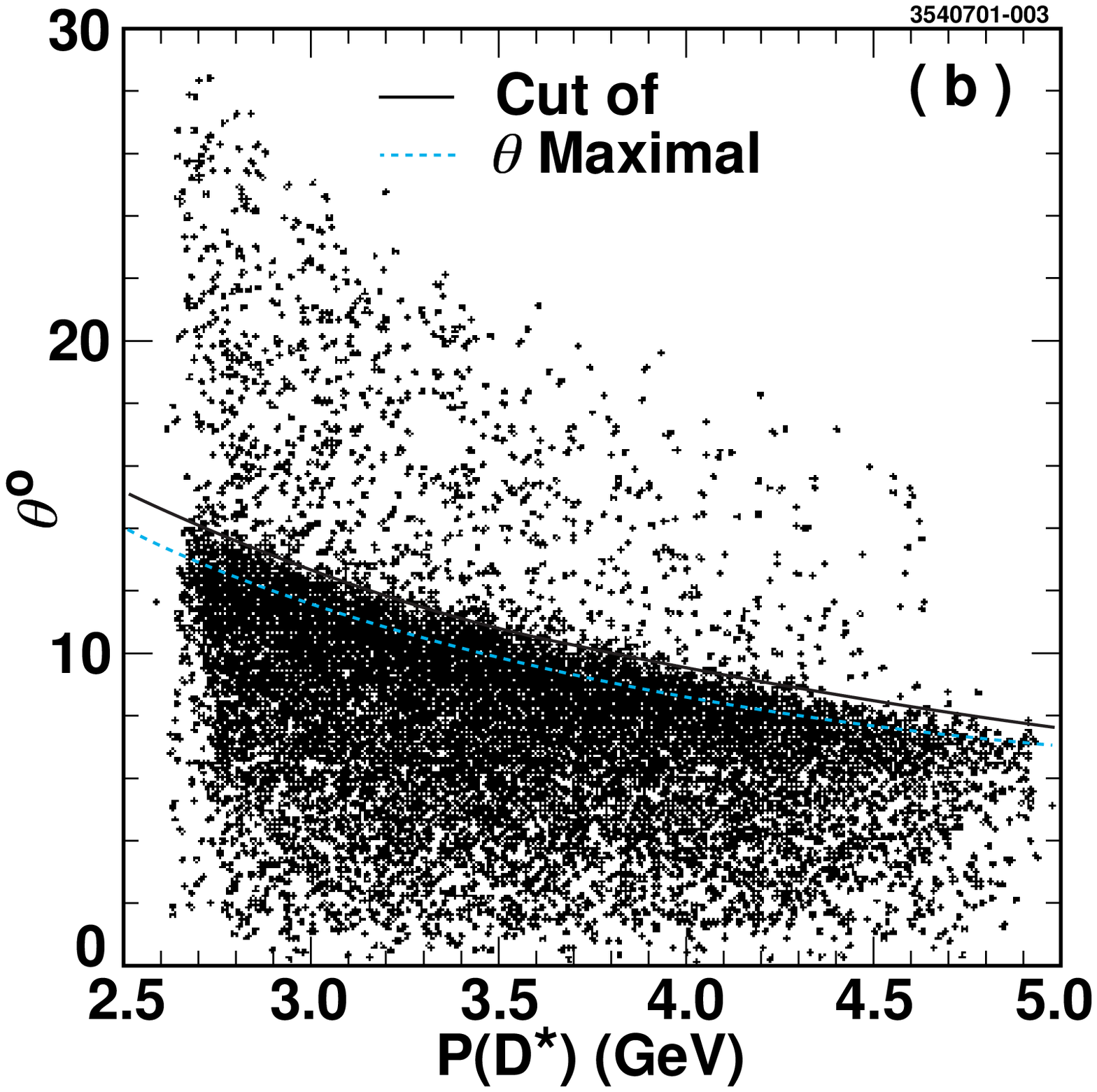}
\end{tabular}
\caption{\label{fig:ppisoft} The $\pi_{\rm slow}^+$ momentum (a)
and the opening angle between $D^0$ and $\pi_{\rm slow}^+$ (b)
both versus the $D^{\ast +}$ momentum in the nominal data sample.}
\end{figure}

	To further improve the quality of reconstruction in our nominal
sample, we
apply some kinematic cuts to remove a small amount of misreconstructed
signal and background.
Figure~\ref{fig:ppisoft} shows the distribution of the momentum
of the $\pi_{\rm slow}^+$ as a function of the $D^{\ast +}$ candidate
momentum.  We apply a cut at the kinematic boundary as shown in the
figure.  Figure~\ref{fig:ppisoft} also shows the opening angle $\theta$ between
the $\pi_{\rm slow}^+$ and the $D^0$ candidate as a function of the
$D^{\ast +}$ candidate momentum.  We apply a cut of
$\theta < 38^\circ/P_{D^\ast}[{\rm GeV}]$ which is just beyond the
kinematic limit to account for resolution smearing.
We also require $\sigma_Q < 200$ keV which removes the long tail in
the error distribution.

	The tracking selected sample makes much more stringent cuts
on the quality of the tracks used to identify the candidates.  All tracks
are required to have hits in both the $r\phi$ and $z$ views
in all three layers of the
silicon strip detector as opposed to the nominal two silicon hits per view.
None of these hits are allowed to be within
2~mm of a silicon wafer edge.  The $D^0$ daughter tracks are required
to have at least 38 of the possible 51 main drift chamber hits
and seven of the ten intermediate drift chamber hits.  The $\chi^2$
per degree of freedom of the fit to these two tracks are limited to
less than 2 in each
of the two drift chambers and 50 in the silicon strip detector.
These selections are designed to remove tracks that have tracking mishaps
or decay in flight.

\begin{figure}
   \begin{minipage}[t]{160mm}
     \epsfxsize=160mm
     \centerline{\epsfbox{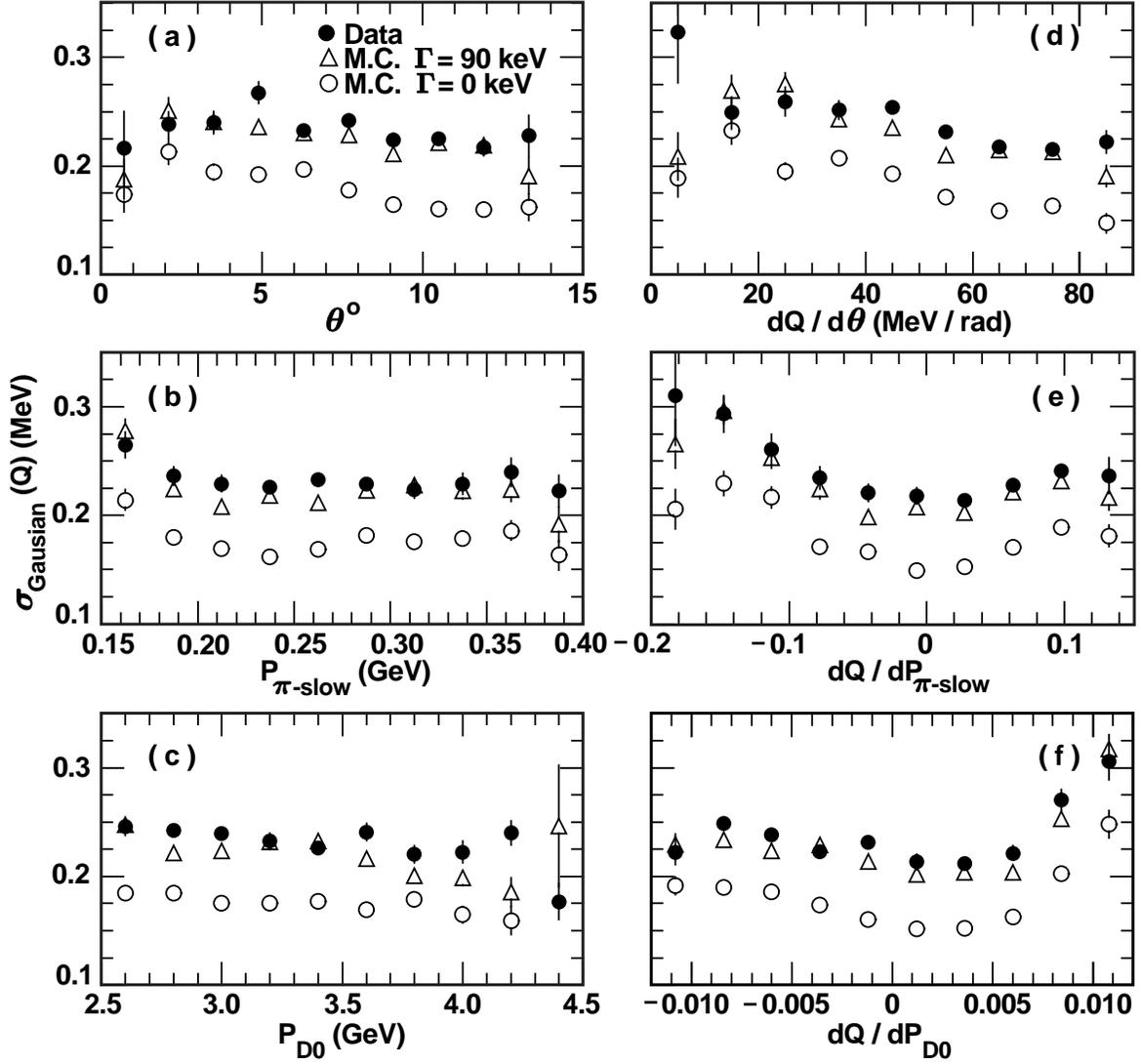}}
     \caption{\label{fig:sigcompare} Gaussian width of $Q$
     distribution versus kinematic parameters and derivatives.
     $\bullet$ -- Data;
     $\circ$--Simulation with $\Gamma_{D^\ast +}=0$;
     $\triangle $--Simulation with $\Gamma_{D^\ast +}=90$ keV. }
   \end{minipage}
\end{figure}

\begin{figure}
   \begin{minipage}[t]{160mm}
     \epsfxsize=160mm
     \centerline{\epsfbox{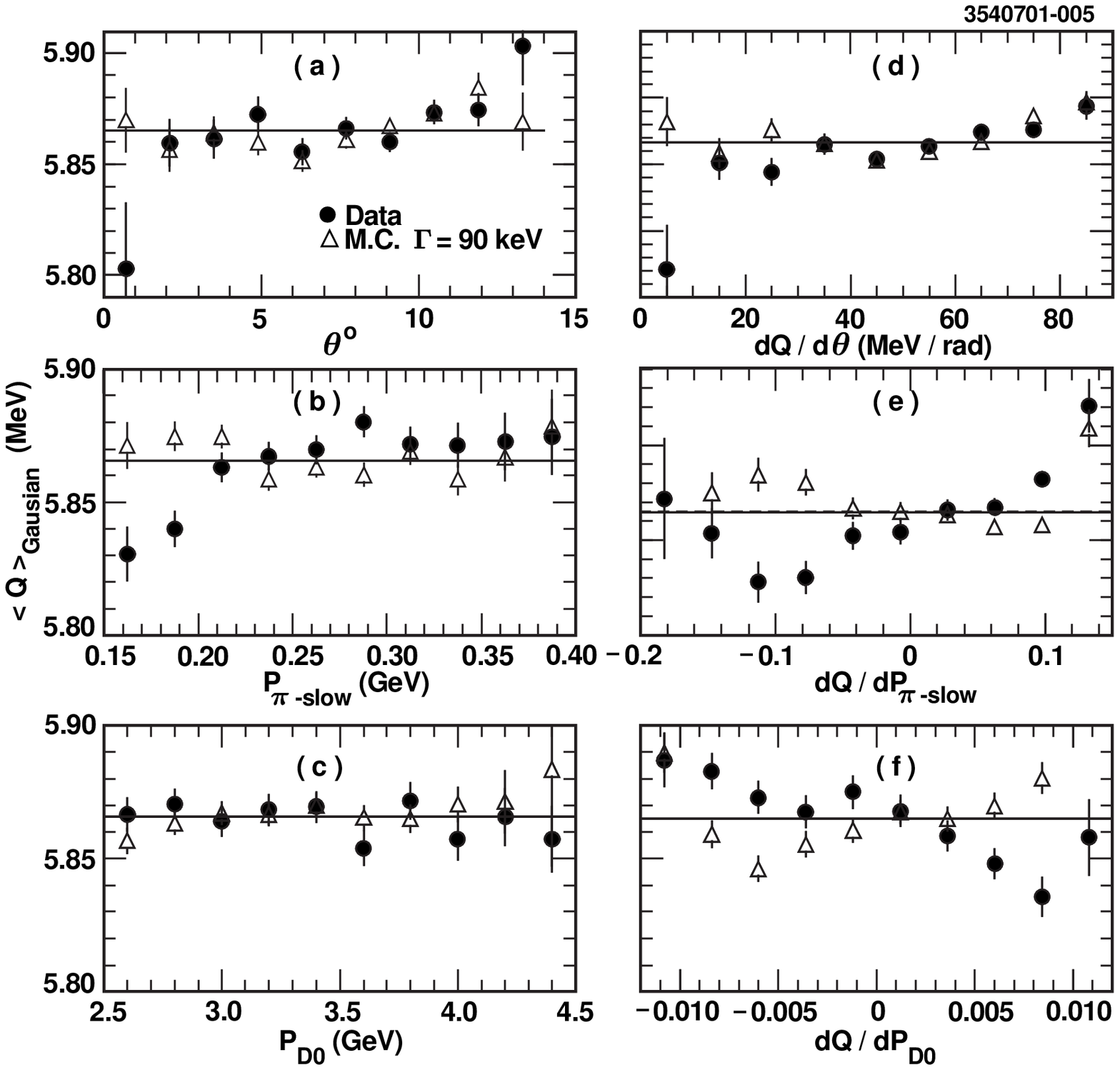}}
     \caption{\label{fig:meancompare} Gaussian mean of $Q$
     distribution versus kinematic parameters and derivatives.
     $\bullet$--Data;
     $\triangle $--Simulation with $\Gamma_{D^\ast +}=90$ keV.
     The horizontal lines show the average value of $Q$ for
     the two samples.  }
   \end{minipage}
\end{figure}

	We compare the simulation and
the data as a function of kinematic variables of the $D^{\ast +}$ decay.
This will provide another test of the simulation's modeling of the data,
and be the basis of our study of systematic uncertainties in the analysis.
The most important kinematic variables are the ``derivatives'' which
are defined by
\begin{equation}
M^2 = M(K\pi)^2+m^2_{\pi^+_{\rm slow}}+ 2(E_{D^0}E_{\pi^+_{\rm 
slow}}-p_{D^0}p_{\pi^+_{\rm slow}}\cos\theta)\ ,
\end{equation}
\begin{equation}
\beta_{D^0} = p_{D^0}/E_{D^0},
\end{equation}
\begin{equation}
\beta_{\pi^+_{\rm slow}} = p_{\pi^+_{\rm slow}}/E_{\pi^+_{\rm slow}},
\end{equation}
\begin{eqnarray}
\frac{\partial Q}{\partial P_{D^0}} & \equiv &
\frac{E_{\pi^+_{\rm slow}}}{M}(\beta_{D^0} - \beta_{\pi^+_{\rm slow}} 
\cos\theta), \\
\frac{\partial Q}{\partial P_{\pi^+_{\rm slow}}} & \equiv &
\frac{E_{D^0}}{M}(\beta_{\pi^+_{\rm slow}} - \beta_{D^0} \cos\theta), \\
\frac{\partial Q}{\partial \theta} & \equiv &
\frac{p_{D^0} p_{\pi^+_{\rm slow}}}{M}\sin\theta.
\end{eqnarray}
These derivatives test correlations among the basic kinematic
variables, the $D^0$ and $\pi^+_{\rm slow}$ momenta and the opening
angle, $\theta$.  We compare by dividing
the $Q$ distribution into ten slices in each of
the kinematic variables and fitting the ten sub-distributions
of $Q$ to Gaussians.  We display the width and mean
of the ten fits as a function of each of the six kinematic
variables in Figures~\ref{fig:sigcompare} and~\ref{fig:meancompare}.

	The quality of the width comparison (Figure~\ref{fig:sigcompare})
is excellent, with the
simulation generated with an underlying $\Gamma(D^{\ast +})$ in
the range of 90--100 keV
agreeing well with the data for all the kinematic variables.  Even when
generated with an underlying $\Gamma(D^{\ast +}) = 0$ keV the simulation
accurately follows the data's changes as the kinematic variables vary
across their allowed range.

	The quality of the mean comparison (Figure~\ref{fig:meancompare})
is not as good.  The dependence
of the mean of $Q$ is not well modeled versus the $\pi^+_{\rm slow}$
momentum, $\partial Q/ \partial P_{\pi^+_{\rm slow}}$, and 
$\partial Q/\partial P_{D^0}$ by our simulation.  We discuss the consequences
of this imperfect modeling of the data in the section on
systematic uncertainties below.

\begin{figure}
\begin{tabular}{cc}
\epsfxsize=85mm \epsfbox{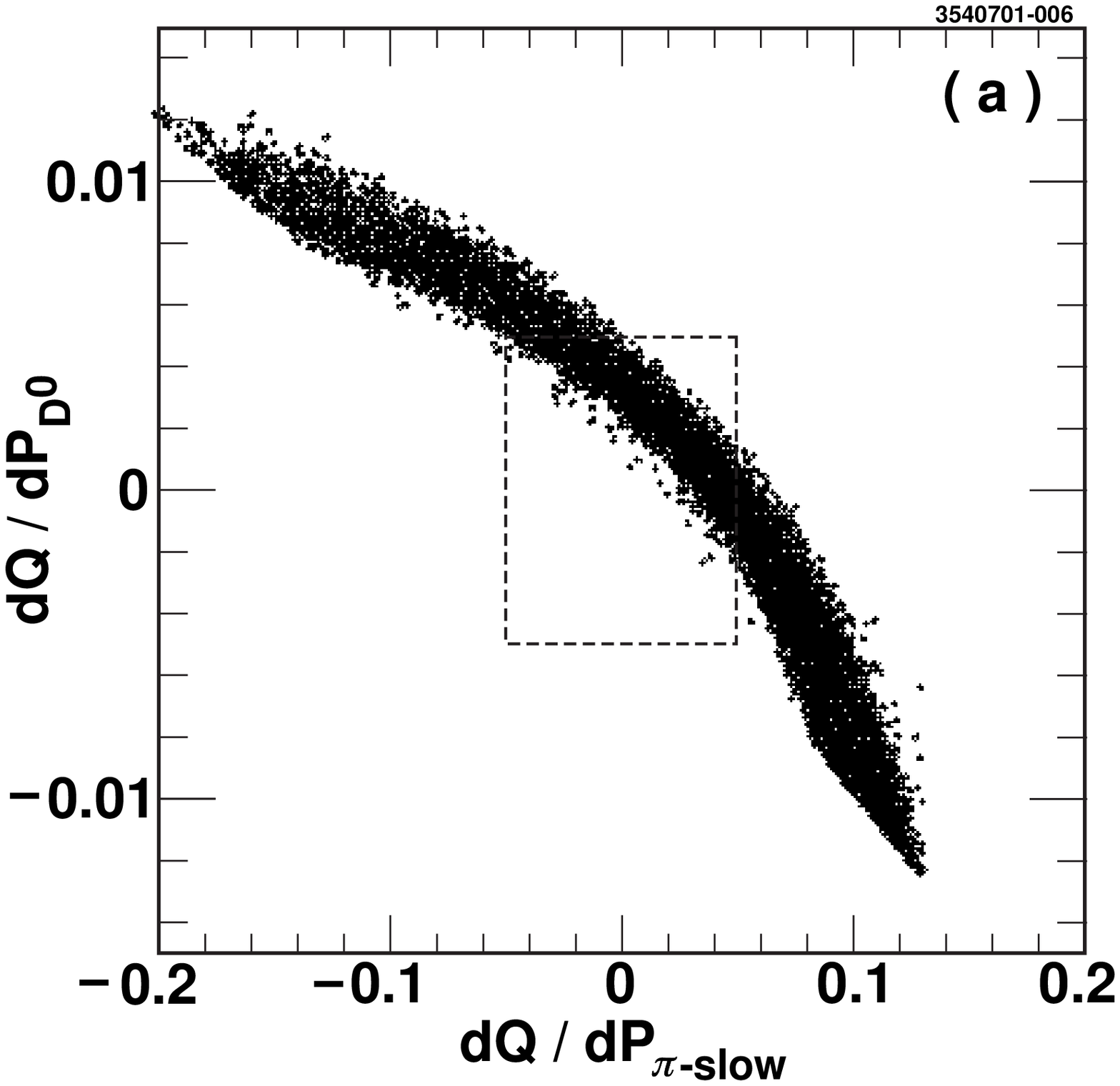} &
\epsfxsize=85mm \epsfbox{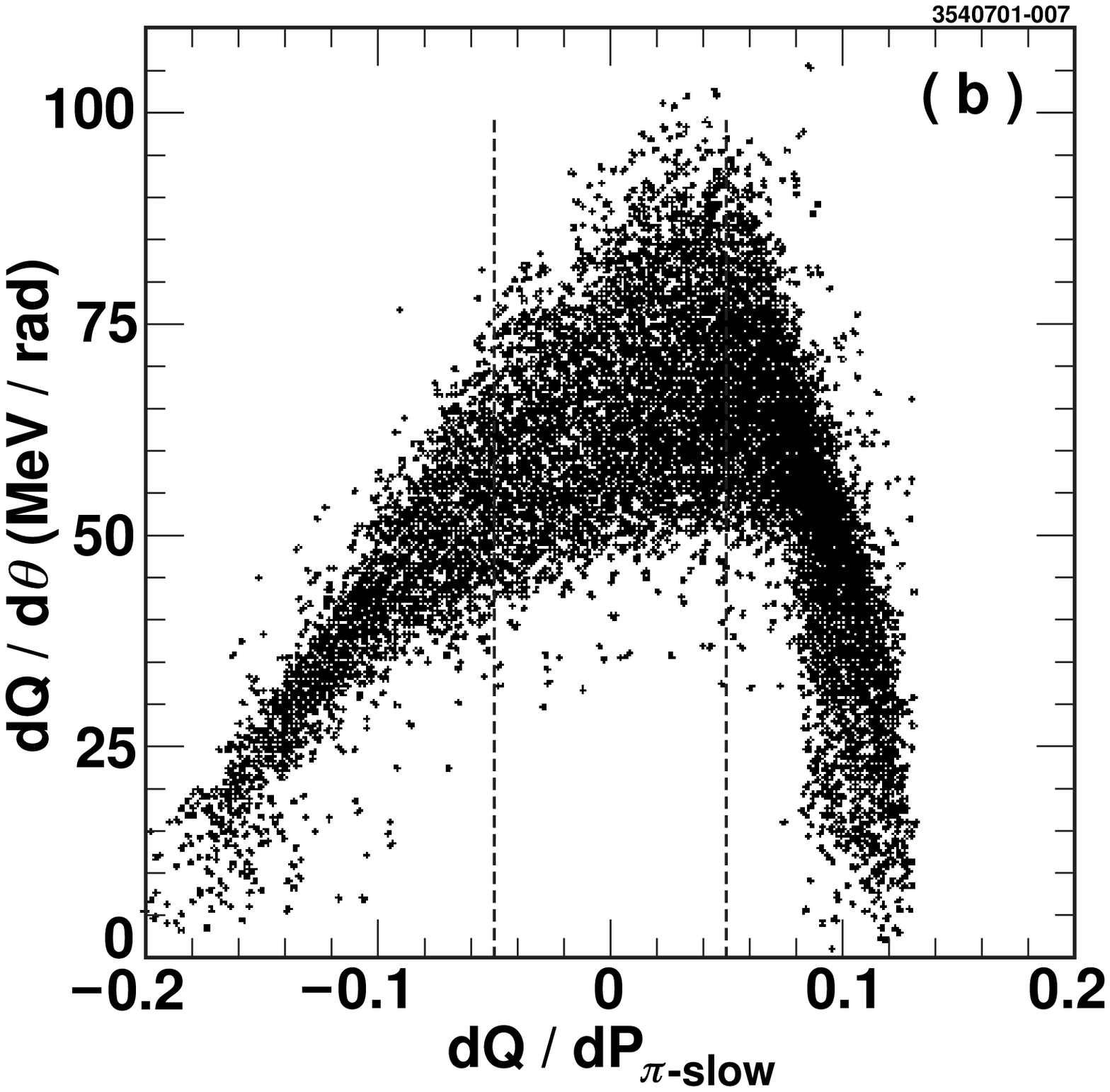} \\
\multicolumn{2}{c} {\epsfxsize=85mm \epsfbox{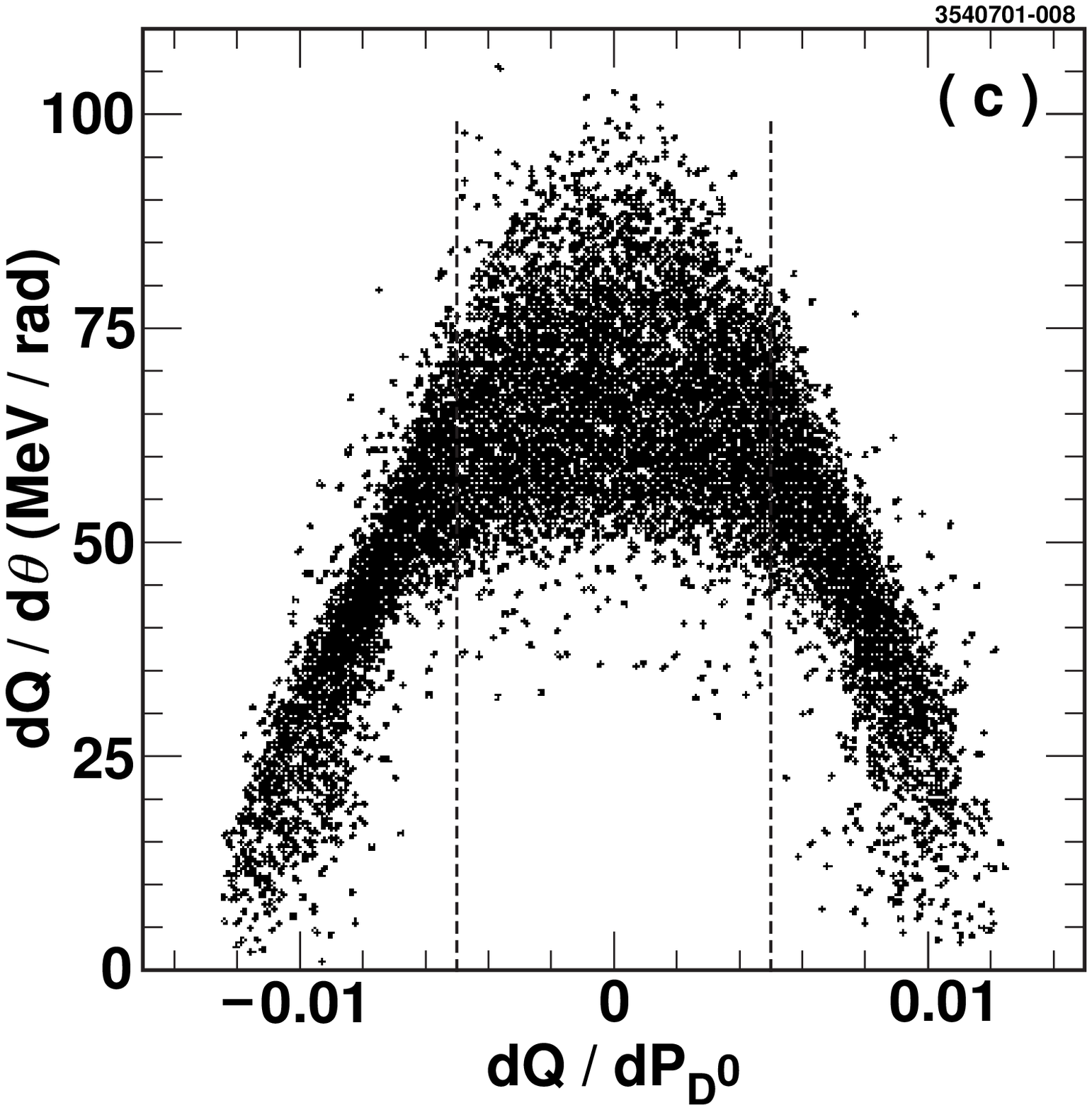}}
\end{tabular}
\caption{\label{fig:derivs} Distributions of $\partial Q /\partial P_{D^0}$
versus $\partial Q / \partial P_{\pi^+_{\rm slow}}$ (a),
$\partial Q /\partial \theta$
versus $\partial Q / \partial P_{\pi^+_{\rm slow}}$ (b), and
$\partial Q /\partial \theta$
versus $\partial Q / \partial P_{D^0}$ (c) in the data.  The dashed
regions show the selection that defines the kinematic selected sample.}
\end{figure}

	Figure~\ref{fig:derivs} shows
the three derivatives plotted against each other in the data.
Note that if we select
$\partial Q/\partial P_{D^0}$ and $\partial Q/\partial P_{\pi^+_{\rm slow}}$
both to be close to zero we minimize the dependence of $Q$ on the
basic kinematic variables $P_{D^0}$ and $P_{\pi^+_{\rm slow}}$,
and thus minimize the contribution of the kinematic variables to the
width of the $Q$ distribution.  With this selection we are more sensitive
to the underlying width of the $Q$ distribution rather than variations caused
by any mismodeling of $Q$'s dependence on the basic kinematics.
The kinematic selection is defined by
\begin{eqnarray}
\left|\frac{\partial Q}{\partial P_{D^0}}\right| & \leq & 0.005, \\
\left|\frac{\partial Q}{\partial P_{\pi^+_{\rm slow}}}\right| & \leq & 0.05.
\end{eqnarray}

	Table~\ref{tab:data} summarizes the statistics in our three samples.
The tracking and kinematic samples are subsets of the nominal sample.
The two subsets contain 94 common candidates.

\section{Fit Description}

	We assume that the intrinsic width of the $D^0$ is negligible,
$\Gamma(D^0) \ll \Gamma(D^{\ast +})$, implying that the width of $Q$
is simply a convolution of the shape given by the $D^{\ast +}$ width and
the tracking system response function.  Thus we
consider the pairs of $Q$ and $\sigma_Q$ for
$D^{\ast +} \to \pi_{\rm slow}^+ D^0 \to K^-\pi^+\pi_{\rm slow}^+$
where $\sigma_Q$ is given for
each candidate by propagating the tracking errors in the kinematic
fit of the charged tracks.  We perform an unbinned maximum likelihood fit
to the $Q$ distribution.

	The underlying signal shape of the $Q$ distribution is assumed to be
given by a P-wave Breit-Wigner
with central value of $Q$, $Q_0$.
We considered a relativistic and non-relativistic
Breit-Wigner as a model of the
underlying signal shape, and found negligible changes in the fit
parameters between the two.
The width of the signal Breit-Wigner
depends on $Q$ and is given by
\begin{equation}
\Gamma(Q) = \Gamma_0  \left(\frac{P}{P_0}\right)^3 \left(\frac{M_0}{M}\right)^2 ,
\label{eq:BW}
\end{equation}
where $\Gamma_0$ is equivalent to $\Gamma(D^{\ast +})$,
$P$ and $M$ are the candidate
$\pi_{\rm slow}^+$ or $D^0$ momentum in the $D^{\ast +}$ rest frame
and $K\pi\pi_{\rm slow}$ mass, and $P_0$ and $M_0$ are
the values computed using $Q_0$.
The effect of the mass term is negligible at our energy. 
The partial width and the total width differ negligibly in  
their dependence on $Q$ for $Q>1~MeV$.

	For each candidate the signal shape is convolved with
a resolution Gaussian with width $\sigma_Q$, determined by the tracking
errors, as a model of our finite resolution shown in
Figure~\ref{fig:errorcompare}.

\begin{figure}
\epsfxsize=\linewidth
\centerline{\epsfbox{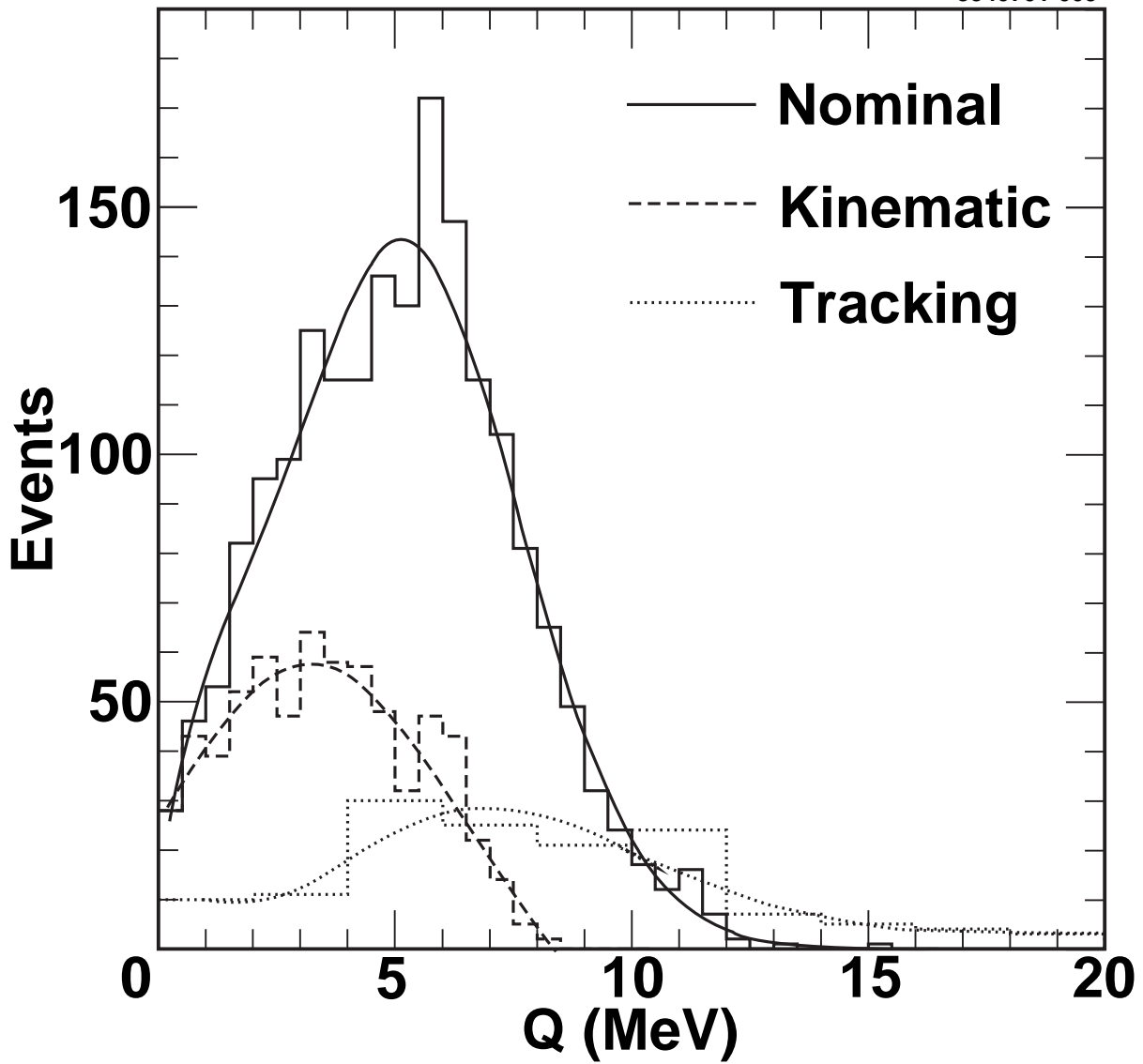}}
\caption{\label{fig:back} Our simulation's prediction of the background
for the three samples discussed in the text.  Also shown are the fits
to third order polynomials that are used in the fits to the data.}
\end{figure}

The fit also includes a background contribution
with a fixed shape.  The shape for the background is taken from fits to the
background prediction of our simulation with a third order polynomial.
The level of the background is allowed to float in our standard
fit.  The predicted background shape and fits are displayed
in Figure~\ref{fig:back}.

\begin{figure}
\epsfxsize=\linewidth
\centerline{\epsfbox{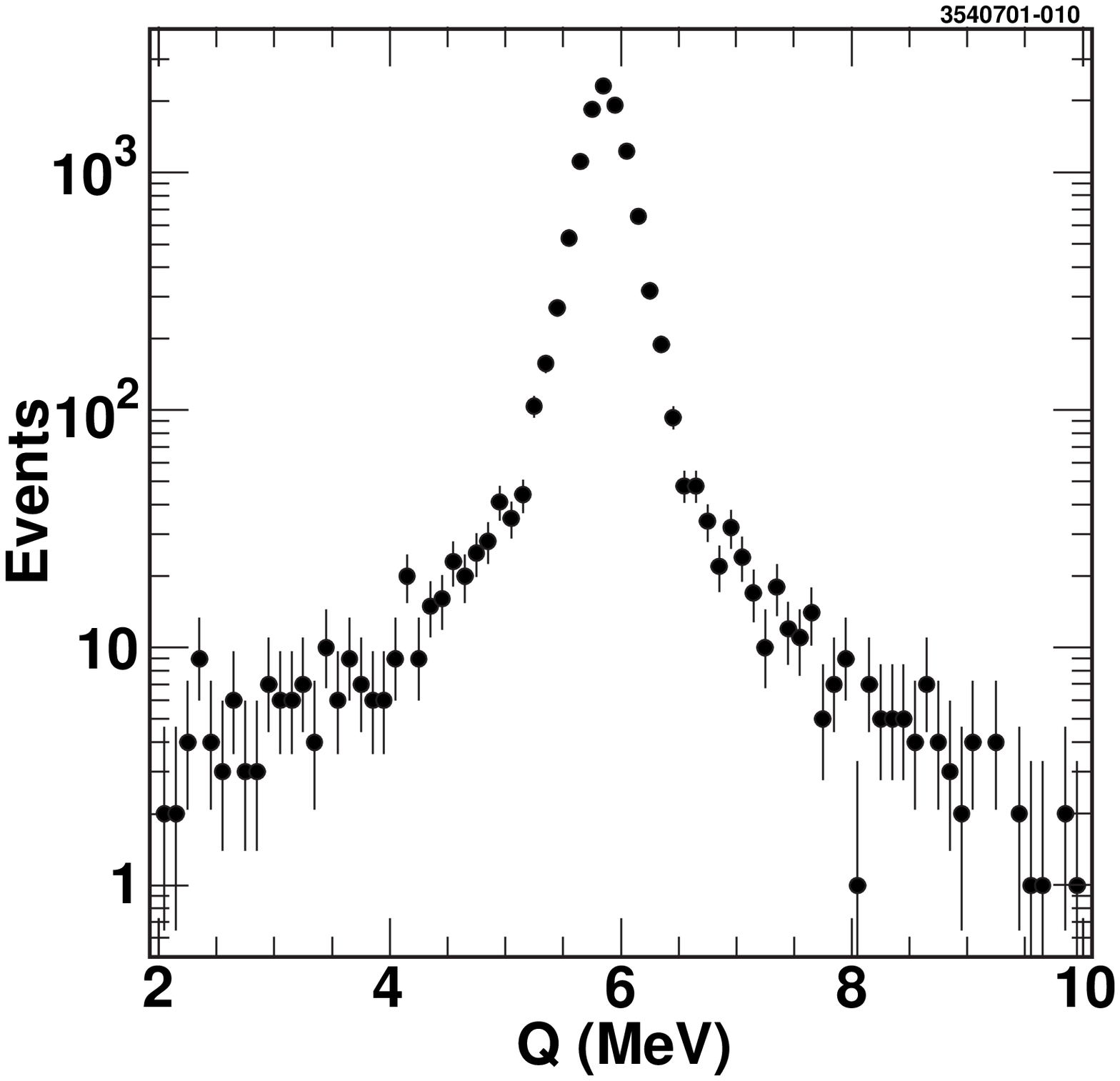}}
\caption{\label{fig:typical} The $Q$ distribution of the nominal
sample in the data.}
\end{figure}

	Figure~\ref{fig:typical} shows the $Q$ distribution for
our nominal data sample.  Note that besides the well measured signal and
the small, slowly varying background, there is also a small
component centered on the signal with a large width.
Therefore we allow a small fraction of the signal, $f_{mis}$,
to be parametrized
by a single Gaussian resolution function of width $\sigma_{mis}$.
This shape is included in the fit to model the tracking
mishaps which our simulation predicts to be at the 5\% level in the
nominal sample and negligible in both the tracking
and kinematic selected samples.  Typically we constrain the level
of this contribution while allowing $\sigma_{mis}$ to float.

	We have many other parameters of the fit that can be varied
or allowed to float for testing purposes.  We can allow a scale factor
on each candidate's $\sigma_Q$ to model a systematic mistake in our
tracking system caused, for
example, by not properly accounting for the material of the detector.
In our standard fits we
only allow the normalization of the background to float, but we can
either vary the shape as indicated by the simulation or
allow the parameters of the background polynomial to float as a measure
of the small systematic uncertainty due to the background shape.

	Table~\ref{tab:parameters} summarizes the parameters of our
\begin{table}
\caption{Parameters of our fit to the $Q$ distribution}
\begin{center}
\begin{tabular}{|c|c|} \hline
Parameter      & Description \\ \hline
$\Gamma_0$     & Breit Wigner width of $Q$ signal distribution, 
$\Gamma(D^{\ast +})$ \\
$Q_0$          & Mean of $Q$ signal distribution \\
$N_s$          & Number of signal events \\
$f_{mis}$      & Fraction of mismeasured signal\\
$\sigma_{mis}$ & Resolution on measured $Q$ for mismeasured signal\\
$N_b$          & Number of background events\\
$k$            & $\sigma_Q$ scale factor, fixed to 1\\
$B_{1,2,3}$    & Coefficients of background polynomial, fixed from simulation\\ \hline
\end{tabular}
\end{center}
\label{tab:parameters}
\end{table}
fit.  Note that the $\sigma_Q$ scale factor $k$ and the background shape
parameters $B_{1,2,3}$ are fixed in our nominal fits.  We minimize the
likelihood function
\begin{equation}
L = 2(N_s + N_b) - 2 \sum _{i=1}^N \log 
[N_s S(Q_i, \sigma_{Qi}; \Gamma_0, Q_0, f_{mis}, \sigma_{mis}, k ) + 
N_b B(Q_i; B_{1,2,3})],
\end{equation}
where $S$ and $B$ are respectively the signal and background shapes
discussed above.

	The fitter has been extensively tested both numerically
and with input from our full simulation.  We find that the fitter
performs reliably giving normal distributions for the floating
parameters and their uncertainties.  It also reproduces the input
$\Gamma(D^{\ast +})$ from 0 to 130 keV.  Its behavior on
each of the three data samples: nominal; tracking selected; and
kinematic selected in the full simulation is discussed below.
We note that if all the parameters are allowed to vary simultaneously
there is strong correlation among the intrinsic width, $\Gamma_0$,
the fraction of mismeasured events, $f_{mis}$, and the $\sigma_Q$ scale
factor, $k$, as one would expect.
Thus our nominal fit holds $k$ fixed, but in our systematic
studies we either fix one of the three or provide a constraint with
a contribution to the likelihood if the parameter varies from its
nominal value.

\section{Fit Results}

	As a preliminary test to fitting the data we run the complete analysis
on a fully simulated sample that has about ten times the data
statistics and is generated with a range of underlying $\Gamma(D^{\ast +})$
from 0~to 130~keV.  We do this for nominal, tracking, and kinematic
selected samples.  For the nominal sample we note that the fit is not
stable if all the parameters are left to vary freely.  We have
found that if we constrain the fraction of mismeasured signal
to $(5.3 \pm 0.5)$\% as indicated by the simulation over the range
of generated widths of the $D^{\ast +}$ then we get a stable result.
This constraint makes the
fit to the simulated nominal sample have no significant offset between
the generated and measured values for the width of the $D^{\ast +}$.
The tracking and kinematic selected samples have a negligible amount
of mismeasured signal according to the simulation and in fits
to these samples we fix $f_{mis}$ to zero.
These simulated samples are also consistent with no offset between
the generated and measured values for the width of the $D^{\ast +}$.
We also note that in all three simulated samples there are no trends
in the difference between measured and generated width as a function
of the generated width; the offset is consistent with zero as a function of
the generated width of the $D^{\ast +}$.
Table~\ref{tab:data} summarizes this simulation
study.  We will apply these offsets to the fit value that we obtain
from the data.  For the energy release all samples show small shifts,
$-7 \pm 3$ keV for the nominal, $-12 \pm 10$ keV for the tracking,
and  $-12 \pm 5$ keV for the kinematic. 

\begin{figure}
\epsfxsize=\linewidth
   \centerline{\epsfbox{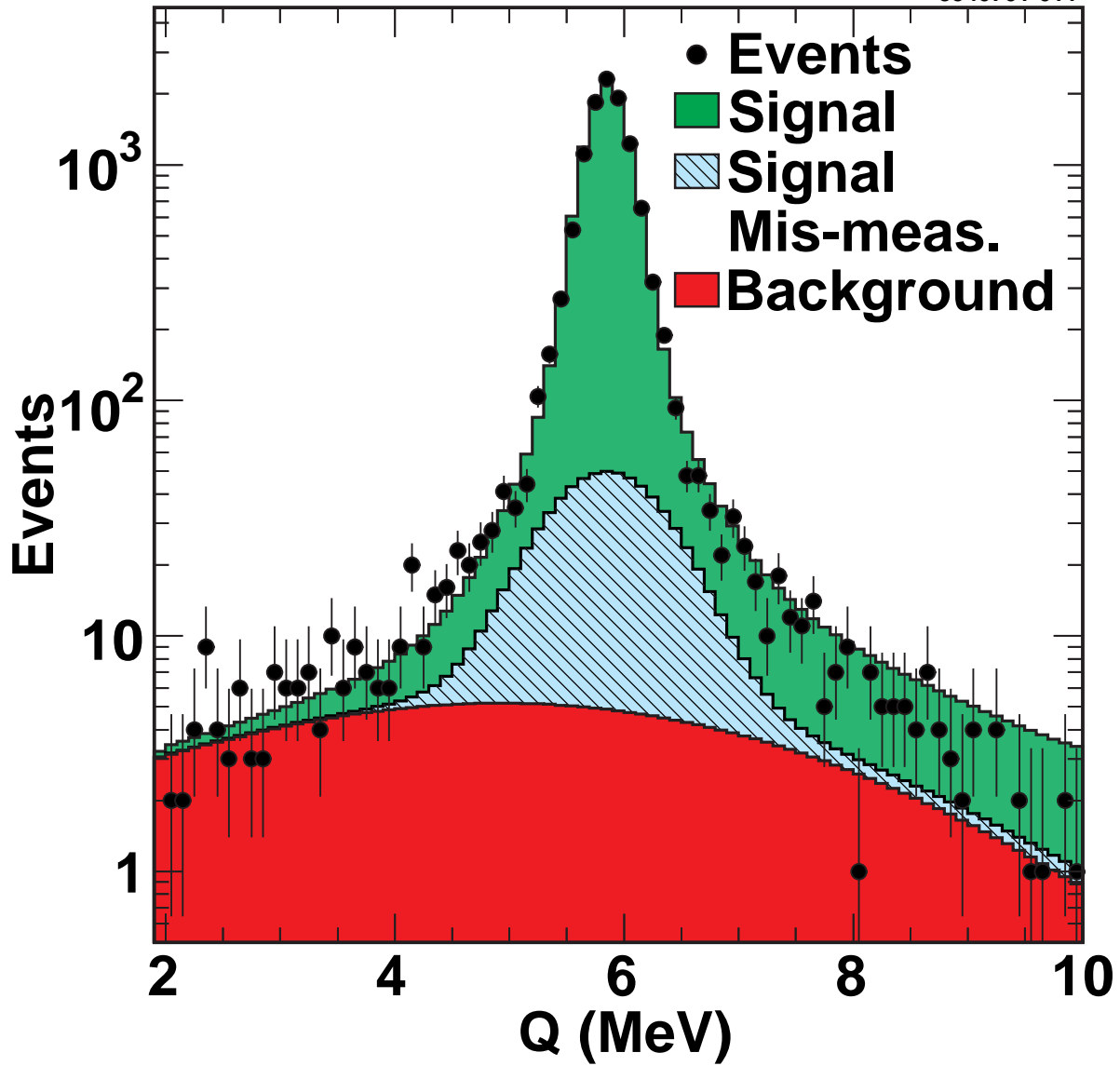}}
   \caption{\label{fig:nomfit} Fit to nominal data sample.  The different
contributions to the fit are shown by different colors.}
\end{figure}

\begin{figure}
\epsfxsize=\linewidth
   \centerline{\epsfbox{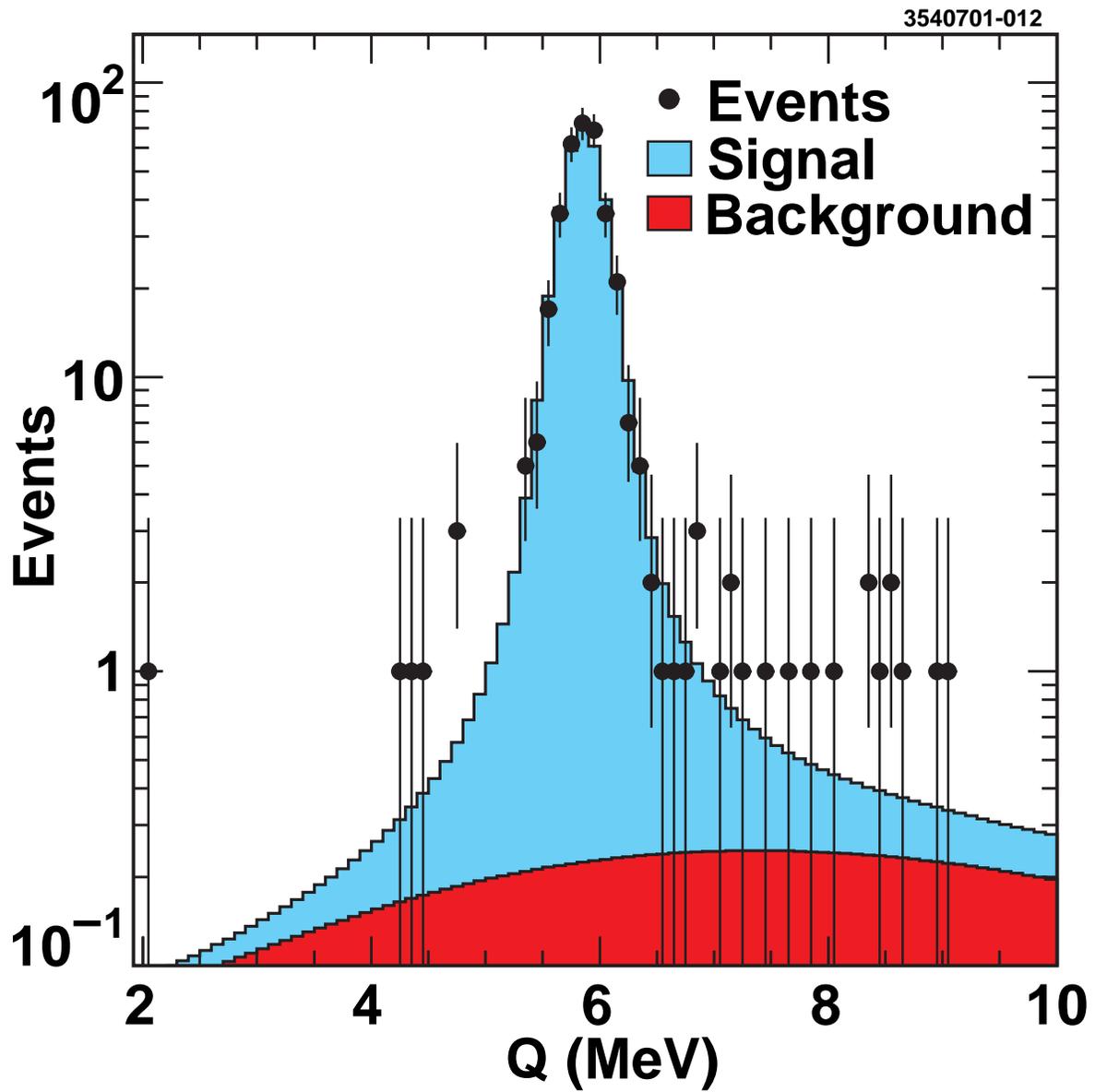}}
   \caption{\label{fig:platinumfit} Fit to tracking selected data sample.
   The different
   contributions to the fit are shown by different colors.}
\end{figure}

\begin{figure}
\epsfxsize=\linewidth
   \centerline{\epsfbox{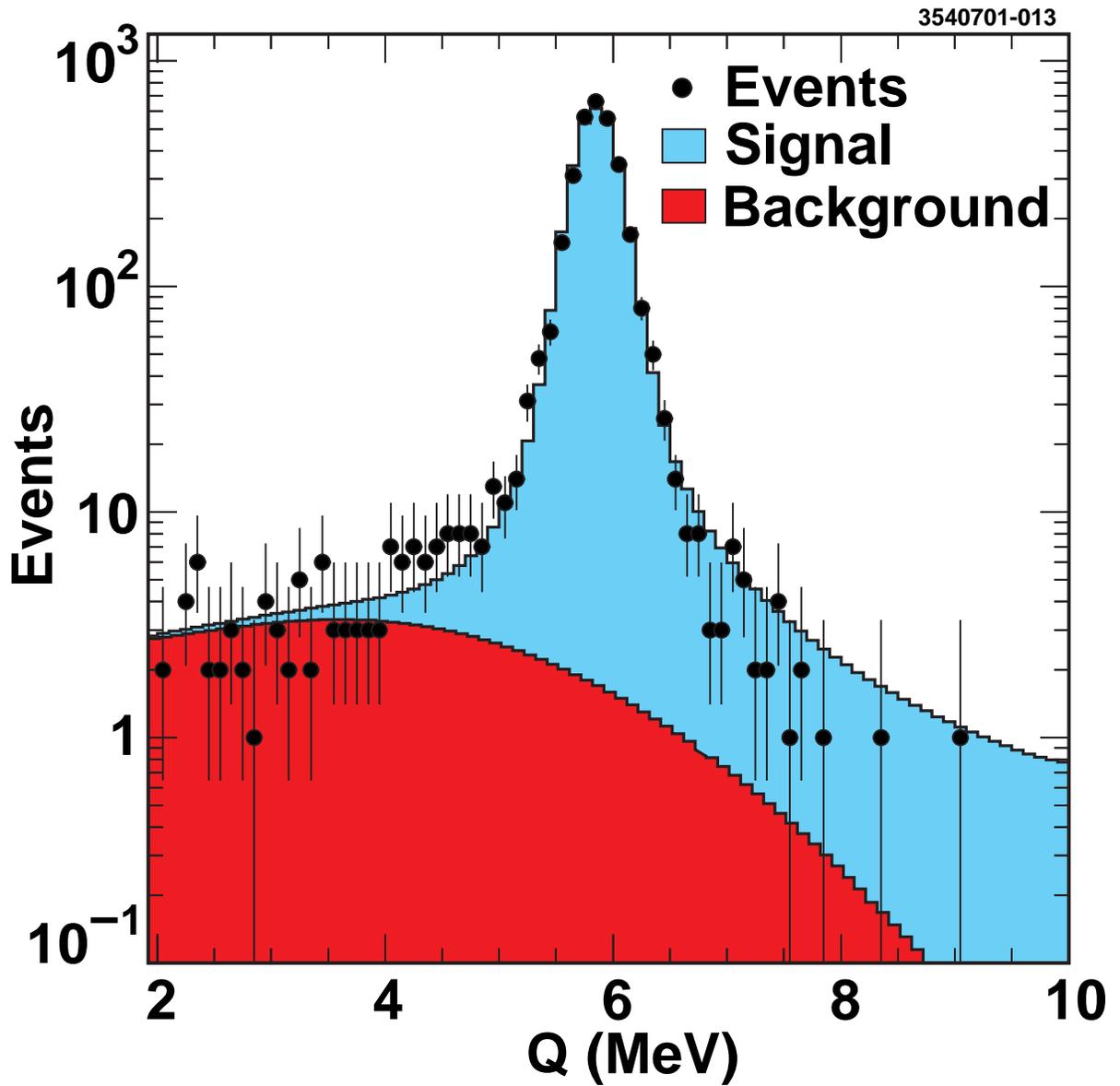}}
   \caption{\label{fig:goldfit} Fit to kinematic selected data sample.
   The different
   contributions to the fit are shown by different colors.}
\end{figure}

\begin{figure}
\begin{tabular}{ccc}
\epsfxsize=55mm
\epsfbox{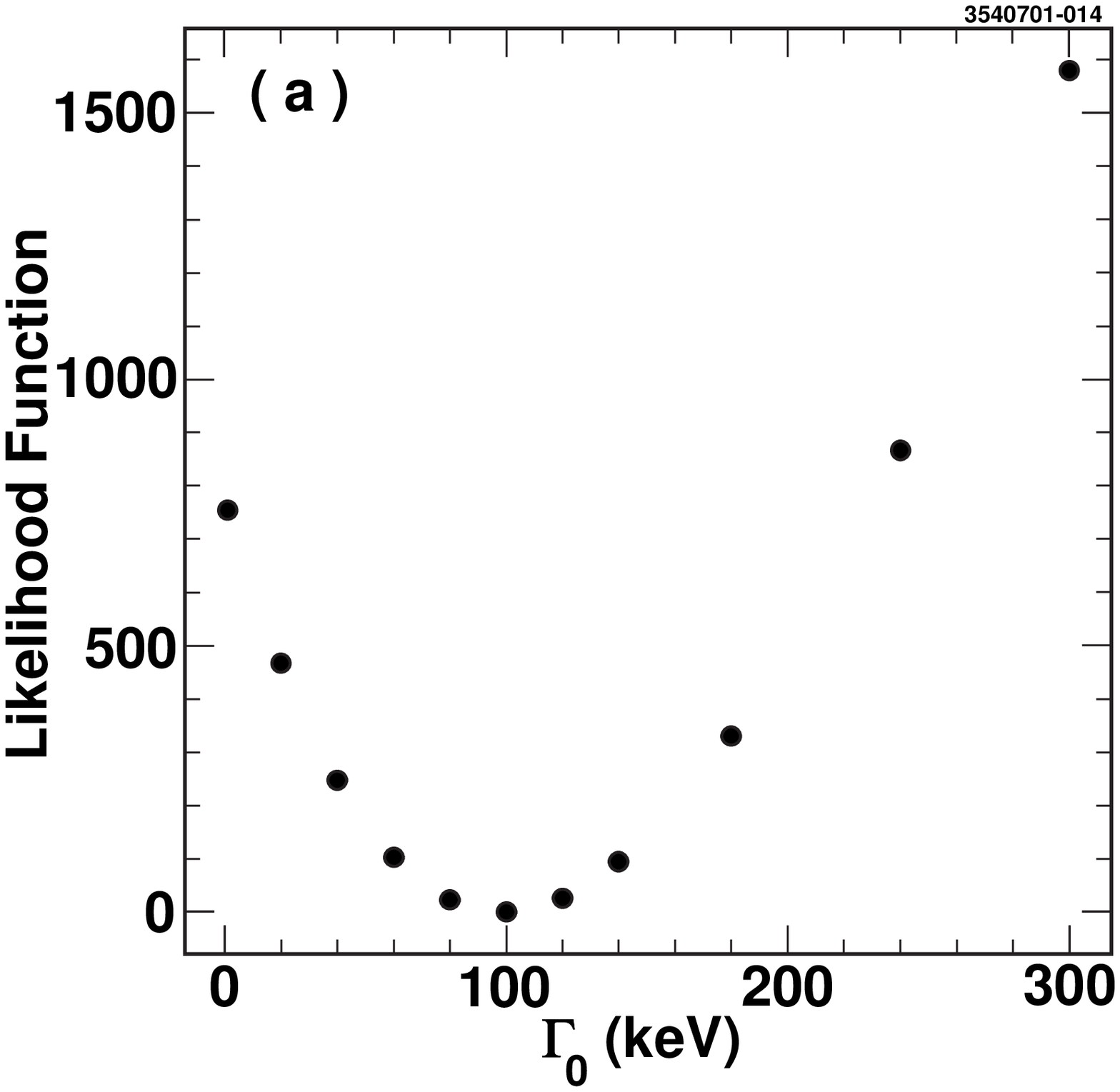} &
\epsfxsize=55mm
\epsfbox{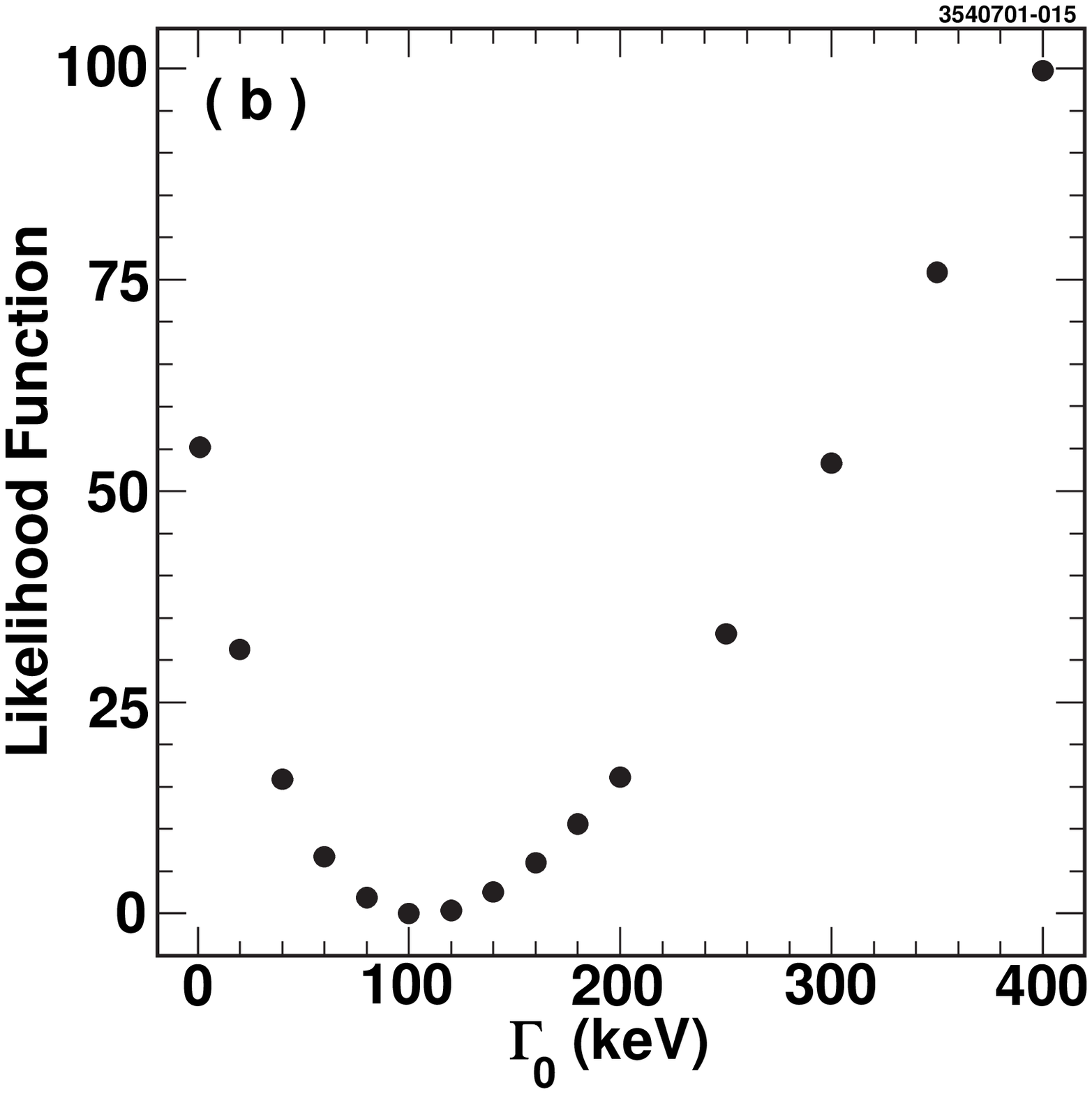} &
\epsfxsize=55mm
\epsfbox{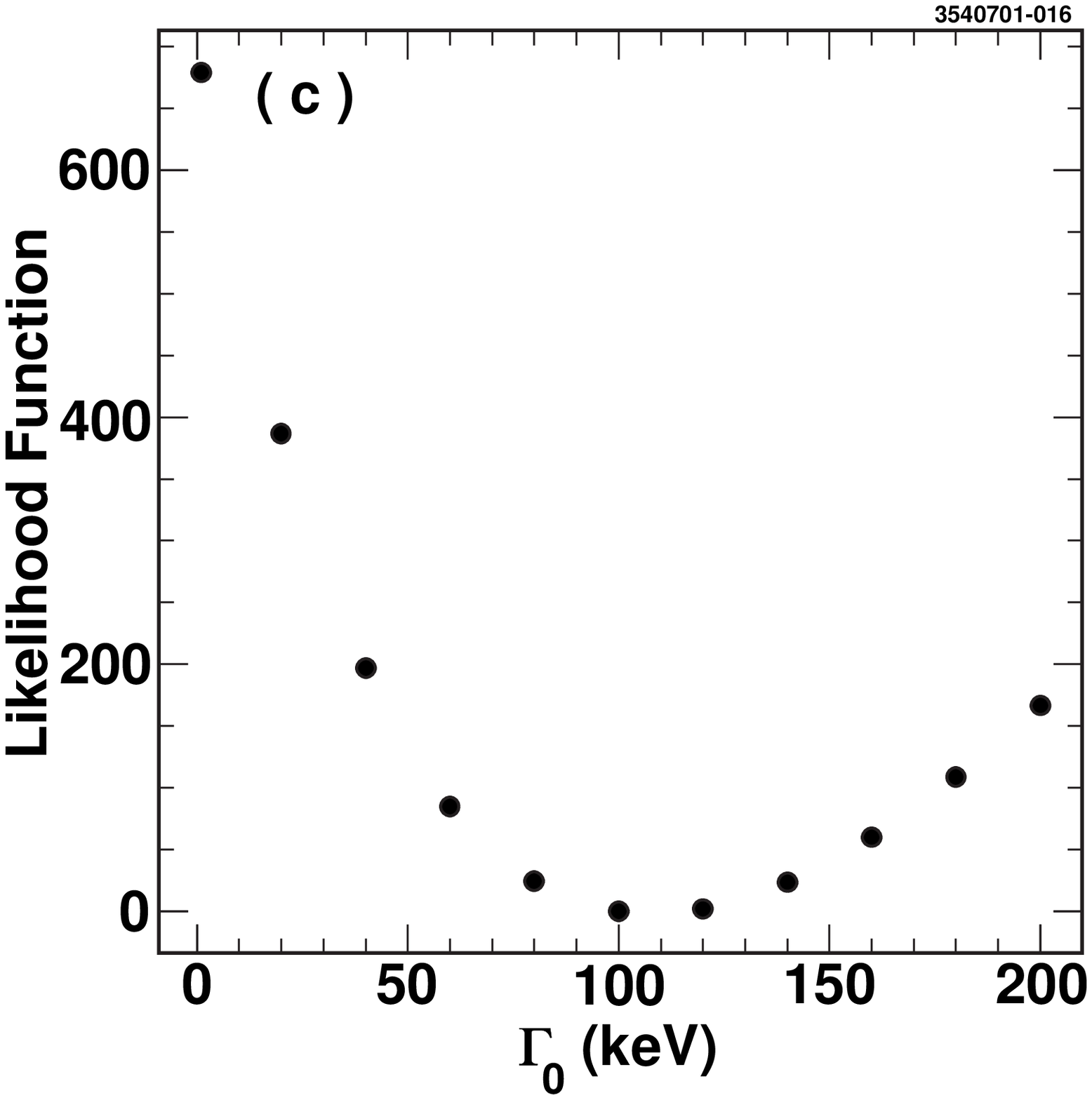} 
\end{tabular}
\caption{\label{fig:like}
Likelihood function versus measured $D^{\ast +}$ width for the nominal (a),
tracking (b), and kinematic (c) selected data samples.}
\end{figure}

	Figures~\ref{fig:nomfit}, \ref{fig:platinumfit}, and \ref{fig:goldfit}
respectively display the fit to the nominal, tracking, and kinematic selected
data samples.  The results of the fits are summarized in Table~\ref{tab:fit}.
\begin{table}
\caption{Results of the fits described in the text.  The fit parameters
are summarized in Table~\ref{tab:parameters}.  The uncertainties are
statistical.}
\begin{center}
\begin{tabular}{|c|c|c|c|} \hline
                      & \multicolumn{3}{c|}{Sample} \\ \hline
Parameter            & Nominal          & Tracking          & 
Kinematic \\ \hline
$\Gamma_0$ (keV)     & $98.9 \pm 4.0$   & $106.0 \pm 19.6$  & $108.1 
\pm  5.9$ \\
$Q_0$ (keV)          & $5853 \pm 2$     & $5854 \pm 10$     & $5850 \pm 4$ \\
$N_s$                & $11207 \pm 109$  & $353 \pm 20$      & $3151 \pm 57$ \\
$f_{mis}$ (\%)       & $5.3 \pm 0.5$    & NA                & NA \\
$\sigma_{mis}$ (keV) & $508 \pm 39 $    & NA                & NA \\
$N_b$                & $289 \pm 31 $    & $15 \pm 7$        & $133 \pm 16$ \\ \hline
\end{tabular}
\end{center}
\label{tab:fit}
\end{table}
Correlations among the floating parameters of the fit are negligible.
Figure~\ref{fig:like}
displays the likelihood as a function of the width of the $D^{\ast +}$
for the fits to the three data samples.

	The agreement is excellent among the three fits, and when the offsets
from Table~\ref{tab:data} are applied we obtain
\begin{eqnarray}
{\rm Nominal\ Sample}     & \Gamma(D^{\ast +}) = 96.2 \pm 4.0\ {\rm keV}, \\
{\rm Tracking\ Selected}  & \Gamma(D^{\ast +}) = 104  \pm 20\ {\rm keV}, {\rm and}\\
{\rm Kinematic\ Selected} & \Gamma(D^{\ast +}) = 103.8 \pm 5.9\ {\rm keV}.
\end{eqnarray}
The data sample and results are summarized in Table~\ref{tab:data}.
\begin{table}
\caption{Summary of the data sample, simulation biases, and fit results.}
\begin{center}
\begin{tabular}{|c|c|c|c|} \hline
                     & \multicolumn{3}{c|}{Sample} \\ \hline
Parameter            & Nominal          & Tracking          & Kinematic \\ \hline
Candidates           & 11496            & 368               & 3284 \\
Background Fraction (\%)
                     & $2.51 \pm 0.27$  & $4.1 \pm 1.9$     & $4.05 \pm 0.49$ \\
$\Gamma_{\rm fit} - \Gamma_{\rm generated}$ (keV)
                     & $2.7 \pm 2.1$    & $1.7 \pm 6.4$     & $4.3 \pm 3.1$ \\
Fit $\Gamma_0$ (keV) &  $98.9 \pm 4.0$  & $106.0 \pm 19.6$ & $108.1 \pm  5.9$ \\       
$D^{\ast +}$ Width (keV) & $96.2 \pm 4.0$
                         & $104 \pm 20$
                         & $103.8 \pm 5.9$ \\ \hline
\end{tabular}
\end{center}
\label{tab:data}
\end{table}
The uncertainties are only statistical.  We discuss systematic
uncertainties in the next section.

\section{Systematic Uncertainties}

	We discuss the sources of systematic uncertainties on our measurements
of the width of the $D^{\ast +}$ in
the order of their size.  The most important contribution is the variation
of the results as a function of the kinematic parameters of
the $D^{\ast +}$ decay as shown 
most clearly in Figure~\ref{fig:meancompare}.  We
estimate this uncertainty by repeating the fits described above  in three
bins for each of the six kinematic parameters and taking the uncertainty as
the largest observed variation from the nominal values in Table~\ref{tab:fit}.
We obtain uncertainties of $\pm 16$ and $\pm 8$ keV on $\Gamma(D^{\ast +})$
and $Q_0$ respectively.

	The next most important contribution
comes from any mismodeling of $\sigma_Q$'s
dependence on the kinematic parameters.  We estimate this by varying
our cut on $\sigma_Q$ from 75 to 400 from the nominal 200 keV and repeating
our analysis with all parameters fixed except allowing the error scale
factor $k$ to vary freely.   This indicates that the resolution is correct
to $\pm 4$\%, and we then repeat our standard analysis with $k$ fixed
at 0.96 and 1.04.  We find uncertainties of $\pm 11$, $\pm 9$,
and $\pm 7$ keV on
$\Gamma(D^{\ast +})$ for the nominal, tracking, and kinematic selected
sample.  For $Q_0$ this uncertainty is negligible except in the tracking
selected sample where it is $\pm 4$ keV.

	We take into account correlations among the
less well measured parameters of the fit, such as $k$, $f_{mis}$, and
$\sigma_{mis}$,
by fixing each parameter at $\pm 1\sigma$ from their central fit values,
repeating the fit, and adding in quadrature the variation in
the width of the $D^{\ast +}$ and $Q_0$ from their central values.
We find uncertainties of $\pm 8$, $\pm 9$, and $\pm 9$ keV on
the width of the $D^{\ast +}$ for the nominal, tracking, and kinematic selected
sample, and respectively $\pm 3$, $\pm 4$, and $\pm 5$ keV on $Q_0$.

	We have studied in the simulation the sources of
mismeasurement that give rise
to the resolution on the width of the $D^{\ast +}$ by replacing the
measured values with the generated values for various kinematic
parameters of the decay products.  We have then compared
these uncertainties with analytic expressions for the uncertainties.
The only source of resolution that we cannot account for in
this way is a small distortion of the kinematics of the event
caused by the algorithm used to reconstruct the $D^0$ origin point
described above.  This contributes an uncertainty $\pm 4$ keV on
the width of the $D^{\ast +}$ and $\pm 2$ keV on $Q_0$.
We have also checked that our simulation accurately models the line shape
of other narrow resonances visible in our data.  Notably the 
decay $\Lambda^0 \to p\pi^-$, has a $Q$ only seven times that of
$D^{\ast+} \to D^0 \pi^+_{\rm slow}$.  In the $\Lambda^0$ decay we
select the $\pi^-$ to have a momentum in the range of those in
the $D^{\ast+}$ decay, and the visible widths agree to a few percent
between data and simulation.

	We consider uncertainties from the background shape by allowing the
coefficients of the background polynomial to float. 
We observe changes on the width of $\pm 4$ keV for the nominal sample
and $\pm 2$ keV for the tracking and kinematic selected samples.
We have also released our kinematic selection cuts which causes the
background to increase by a large factor.  This causes a change which
is small compared to allowing the coefficients of the background shape
polynomial to float.  Variations in the background have a negligible
effect on $Q_0$.
   
	Minor sources of uncertainty are from the width offsets derived
from our simulation and given in Table~\ref{tab:data}, and
our digitized data storage format which saves track parameters
with a resolution of 1~keV and contributes an uncertainty
of $\pm 1$ keV on the width of the $D^{\ast +}$ and $Q_0$.

	An extra and dominant source of uncertainty on $Q_0$ is the energy
scale of our measurements.  We evaluate this uncertainty by selecting
$K_s \to \pi^+\pi^-$ decays in our data.  The daughters tracks of
the $K_s$ candidates are required to pass the same selection
criterion as those described above in the nominal sample, the
decay vertex is required to be inside the beam pipe, and the
vertex is required to be significantly separated from the 
overall event vertex.  Our $K_s$ sample is quite clean, less
than 1\% background under the mass peak, and has millions of candidates.
We then plot the
mean of the $\pi^+\pi^-$ invariant mass as function of the momentum
of the daughters.  We find that above a daughter momentum of 500 MeV/c 
the $K_s$ mass agrees with its expected value~\cite{PDG}.  We apply
corrections, less
than 0.3\% relative, to tracks between 100 and 500 MeV/c to bring the mass
peaks into agreement with the nominal value.  These corrections only 
affect the slow pion and produce a shift in $Q_0$
of $-4$~keV and a negligible change in the width.  We evaluate uncertainties
in the energy scale by varying an overall momentum scale to give
a $\pm30$ keV variation, the uncertainty, of the $K_s \to \pi^+\pi^-$ mass,
and applying the statistical errors we obtain
from the calculations of the momentum corrections discussed above.
Conservatively we add in quadrature twice the observed shift.  We observe
an uncertainty of 8 keV on $Q_0$ and 1 keV on the width due to uncertainty
in the energy scale of our measurements.

	Table~\ref{tab:systematic} summarizes the systematic
\begin{table}
\caption{Systematic uncertainties on the width of the $D^{\ast +}$ and $Q_0$}
\begin{center}
\begin{tabular}{|c|c|c|c|c|c|c|} \hline
                         & \multicolumn{6}{c|}{Uncertainties in keV} \\
                         & \multicolumn{6}{c|}{Sample} \\ \hline
                         & \multicolumn{2}{c|}{Nominal}
                         & \multicolumn{2}{c|}{Tracking}
                         & \multicolumn{2}{c|}{Kinematic} \\ \hline

Source                    & $\delta \Gamma(D^{\ast +})$ & $\delta Q_0$
                           & $\delta \Gamma(D^{\ast +})$ & $\delta Q_0$
                           & $\delta \Gamma(D^{\ast +})$ & $\delta Q_0$ \\ \hline
Dependence on Kinematics  & 16 &  8   & 16 &  8   & 16 &  8 \\
Mismodeling of $\sigma_Q$ & 11 & $<1$ &  9 &  4   &  7 & $<1$ \\
Fit Correlations          &  8 &  3   &  9 &  4   &  9 &  5 \\
Vertex Reconstruction     &  4 &  2   &  4 &  2   &  4 &  2 \\
Background Shape          &  4 & $<1$ &  2 & $<1$ &  2 & $<1$ \\
Offset Correction         &  2 &  3   &  6 & 10   &  3 &  5 \\
Energy Scale              &  1 &  8   &  1 &  8   &  1 &  8 \\
Data Digitization         &  1 &  1   &  1 &  1   &  1 &  1 \\ \hline
Quadratic Sum             & 22 & 12   & 22 & 16   & 20 & 14 \\ \hline
\end{tabular}
\end{center}
\label{tab:systematic}
\end{table}
uncertainties on the width of the $D^{\ast +}$ and $Q_0$.


\section{Conclusion}

	We have measured the width of the $D^{\ast +}$ by studying the
distribution of the energy release in $D^{\ast +} \to D^0 \pi^+$ followed
by $D^0 \to K^- \pi^+$ decay.  We have done this in three separate samples,
one that is minimally selected, a second that reduces poorly
measured tracks
due to misassociated hits and non-Gaussian scatters in the detector
material, and a third that takes advantage of the kinematics of the decay
chain to reduce dependence on mismeasurements of kinematic parameters.
The resolution on the energy release is well modeled by our simulation,
with agreement between the sources of the resolution as predicted
by the simulation and analytic calculations.  
The largest sources of uncertainty are
imperfect modeling of the dependence of the mean energy release on the
kinematics of the decay chain, 
the simulation of the error on the energy release,
and correlations among the parameters of the fit to the energy release
distribution.  
With our estimate of the systematic uncertainties for each of the
three samples being essentially the same we chose to report the result
for the sample with the smallest statistical uncertainty, the minimally
selected sample, and obtain
\begin{equation}
\Gamma(D^{\ast +}) = 96 \pm 4 \pm 22\ {\rm keV},
\label{eq:result}
\end{equation}
where the first uncertainty is statistical and the second is systematic.
We note that if we form an average value taking into account the
statistical correlations among our three measures we get a result
that is nearly identical with Equation~\ref{eq:result} since
the average is dominated by the result with the smallest statistical
uncertainty.  

This is the first measurement of the width of
the $D^{\ast +}$, and our
measurement corresponds to a strong coupling\cite{pred}
\begin{equation}
g = 0.59 \pm 0.01 \pm 0.07,
\end{equation}
and
\begin{equation}
g_{D^\ast \to D\pi} = 17.9 \pm 0.3 \pm 1.9.
\end{equation}
This is consistent with theoretical predictions based on HQET and
relativistic quark models, but higher than predictions based on QCD
sum rules.  

	We also measure the mean value for the energy release in
$D^{\ast +} \to D^0 \pi^+$ decay
\begin{equation}
Q_0 = 5842 \pm 2 \pm 12\ {\rm keV},
\end{equation}
where the first error is statistical and second is systematic.
Combining this with the mass of the charged pion, 139.570 MeV, with
an uncertainty less than 1 keV \cite{PDG}, we calculate
\begin{equation}
m_{D^\ast(2010)^+} - m_{D^0} = 145.412 \pm 0.002 \pm 0.012\ {\mathrm MeV}.
\end{equation}
This agrees with the value from the Particle Data Group,
$145.436 \pm 0.016$ MeV,
from a global fit of all flavors of $D^\ast$--$D$ mass differences.
It also agrees well with the best previous measure from a single experiment
that includes an evaluation of systematic uncertainties
from ACCMOR at $145.39 \pm 0.07$ MeV~\cite{ACCMOR}.

\section*{Acknowledgments}

We thank D.~Becirevic,
I.~I.~Bigi, G.~Burdman, A.~Khodjamirian, P.~Singer, and A.~L.~Yaouanc
for valuable discussions.
We gratefully acknowledge the effort of the CESR staff in providing us with
excellent luminosity and running conditions.
M. Selen thanks the PFF program of the NSF and the Research Corporation, 
and A.H. Mahmood thanks the Texas Advanced Research Program.
This work was supported by the National Science Foundation, the
U.S. Department of Energy, and the Natural Sciences and Engineering Research 
Council of Canada.


\begin{thebibliography}{99}

\bibitem{pred} V.M.Belyaev {\it et al.} Phys. Rev. D 51, 6177 (1995)
               contains a recent survey summarizing and referencing previous
               theoretical work.  
               P. Singer Acta Phys. Polon. B30 3849 (1999),  
               J. L. Goity and W. Roberts JLAB-THY-00-45 (hep-ph/0012314),
               K. O. E. Henriksson, {\it et al.} Nuc. Phys. A 686, 355 (2001),
               and M. Di Pierro and E. Eichten hep-ph/0104208
               appear since that survey.
\bibitem{wise} M. Wise, Phys. Rev. D {\bf 45}, R2188 (1992),
               G.~Burdman and J.~F.~Donoghue,
               Phys.\ Lett.\ B {\bf 280}, 287 (1992), and 
               T.~Yan {\em et al.},
               Phys.\ Rev.\ D {\bf 46}, 1148 (1992)
               [Erratum-ibid.\ D {\bf 55}, 5851 (1992)].
               D.~Becirevic and A.~L.~Yaouanc,
               JHEP {\bf 9903}, 021 (1999) contains
               a recent review referencing previous theoretical work.
\bibitem{mats} J.Bartelt {\em et al.} (CLEO Collaboration),
               Phys. Rev. Lett. {\bf 80}, 3919, (1998).
\bibitem{burdman} G.~Burdman, Z.~Ligeti, M.~Neubert and Y.~Nir,
               Phys.\ Rev.\ D {\bf 49}, 2331 (1994).
\bibitem{ACCMOR} S.Barlag {\em et al.},
                 Phys. Lett. {\bf B 278}, 480 (1992).
\bibitem{GEANT} R. Brun {\it et al.}, GEANT3 Users Guide, CERN DD/EE/84-1.
\bibitem{CLEO} Y.~Kubota {\it et al.}, (CLEO Collaboration),
               Nucl. Instrum. Methods Phys. Res., Sect. A {\bf 320},
               66 (1992); 
               T.~Hill, Nucl. Instrum. Methods Phys. Res.,
               Sect. A {\bf 418}, 32 (1998).
\bibitem{kalman} P. Billior, Nucl. Instrum. Methods Phys. Res.,
               Sect. A {\bf 225}, 352 (1984).
\bibitem{Dmix} R. Godang {\it et al.} (CLEO Collaboration), Phys.
                Rev. Lett. {\bf 84},5038 (2000)
\bibitem{hourglass} D. Cinabro {\it et al.}, CLNS 00/1706, physics/0011075.
\bibitem{PDG} D. E. Groom {\it et al.} (Particle Data Group),
              Eur.Phys.J. {\bf C 15}, 1 (2000).

\end{thebibliography}
\end{document}